\documentstyle[ltwol]{article}

\def\be{\begin{equation}}
\def\ee{\end{equation}}
\def\bea{\begin{eqnarray}}
\def\eea{\end{eqnarray}}

\begin{document}

\title{PARALLEL COMPUTING FOR QCD ON A PENTIUM CLUSTER}

\author{X. Q. LUO, E. B. GREGORY}
\address{Department of Physics, Zhongshan University, Guangzhou
510275, People's Republic of China\\E-mail: stslxq@zsu.edu.cn, 
steric@zsu.edu.cn}   
\author{J. C. YANG, Y. L. WANG, D. CHANG, and Y. LIN}

\address{Guoxun, Ltd, Guangzhou, People's Republic of China}
\twocolumn[\maketitle\abstracts{Motivated by the computational demands of 
our research and budgetary constraints which are common to many research
institutions,
we built a ``poor man's supercomputer'', a cluster of PC nodes which together 
can perform parallel calculations at a fraction of the price of a commercial
supercomputer.  We describe the construction, cost, and performance of our 
cluster.}]

\section{Introduction}
The lattice field theory group at the Zhongshan University physics department
is in a period of rapid development. The group's interests involve such topics
as finite density QCD, Hamiltonian Monte Carlo lattice field theory, lattice
supersymmetry, and lattice quantum gravity. All of these topics can be 
studied through Monte Carlo simulation, but can be quite costly in 
terms of computing power.  In order to do large scale 
numerical investigations of these topics, we required a corresponding 
development of our local computing resources.  We decided building a PC 
cluster capable of parallel computing was the most economical way to build 
computing power.  This type of cluster is often termed a ``Beowulf Cluster'' 
and was pioneered by the United States' National Aeronautics and Space 
Administration. The Beowulf Project's website \cite{beowulf} also proved to be a valuable resource 
in the construction of our cluster.

\section{Construction}
\subsection{Hardware}
One big advantage of a PC cluster over other types of supercomputers is the 
low cost and easy availability of the hardware components. All the hardware 
in our cluster is available at retail computer suppliers. This gives us great 
flexibility in both building the cluster and in any future upgrades or 
expansions we may choose to make. 

At the present our cluster consists of ten PCs, each one has two 500MHz 
Pentium III processors. Additionally, each has 128MB of RAM, an 8GB EIDE
hard disk,
a 100Mbit/sec Ethernet card, a CDROM, a floppy drive and a basic graphics card.
In practice the CDROM, the floppy drive, and even the graphics card could be 
considered extraneous,  as all interaction with the nodes could be done 
through the network.  However, with these 
components, all of which are relatively cheap in comparison to the total cost, 
the operating system installation and occasional maintenance is 
significantly easier.  One node has 
a larger hard disk (20GB) and a SCSI adaptor, for communication with a tape 
drive for disk backups.  The entire cluster shares one monitor, mouse and 
keyboard.

A 100 Mbit/s fast Ethernet switch handles the inter-node communication.  The 
switch has 24 ports so the cluster is expandable to a total of 48 processors 
using the current scheme.  Of course it is possible to link multiple switches
or use nodes with more that two processors, so the possibilities for a larger 
cluster are nearly boundless.
\subsection{Software}
The cluster runs on the Linux operating system.  The reasons for choosing Linux 
are manifold. It is cheap. It can easily support multiple users. It easily
supports network file systems (shared hard disks) and allows user accounts to
be shared across the cluster. Furthermore, compilers for C, C++, and Fortran 
are free.  

To operate the cluster as a parallel computer, the programmer must design the 
algorithm so that it appropriately divides the task among the individual 
processors. He or she must then include appropriate message passing 
functions in the code which allow information to be sent and received by the 
various processors.  We use MPI (Message Passing Interface) \cite{mpi} one of the most
popular message passing standards.  By adopting such a widely used standard
 we are able to share C, C++, and Fortran programs with other colleagues who
may be using any of a large variety of computing platforms.

\section{Performance and Cost}
\subsection{General Cost Comparison}
The LINPACK \cite{linpack} benchmarking test on a single processor shows that a
single 500MHz Pentium processor has a peak speed of about 100 Mflops (100 
million floating point operations per second).  The peak speed for a cluster 
of 20 such processors therefore approaches 2 Gflops.

The cost of our cluster was a little under US\$14000 in 1999.  This gives an 
approximate cost of US\$7 per Mflop.  For comparison we can examine the 
cost of 
a commercially produced supercomputer.  The Cray company offers a staring model
of its T3E-1200E supercomputer cluster for US\$630000 \cite{cray}. This 
includes six
processors, each capable of processing at a peak speed of 1200 Mflops.  The 
cost per Mflop on the T3E-1200E machine is therefore US\$87.50.  We have 
constructed a machine that is an order of magnitude cheaper per Mflop. 

We should note here that six of our two-processor nodes are not 
equivalent to a single Cray node.  One of the slowest parts of a parallel 
computation is the communication.  Fewer faster nodes always require less 
communication and the communication channels are inherently faster on the 
Cray machine. 
However, some problems are by their nature easier to divide into nearly 
independent parts.  These problems require less communication and the 
penalty for  a system with slower communication or more nodes is smaller.
It is for just these types of problems that a PC cluster such as ours is
particularly well suited.  One can take full advantage of the processing power.
For most  easily parallizable problems, a PC cluster seems to the most 
economical solution for way of providing computing power.

\subsection{QCD Benchmarking}
In particular, we are interested in using for Monte Carlo simulations of 
lattice quantum chromodynamics (QCD).  Lattice QCD simulations are well suited
for parallelization \cite{gupta} as they involve mostly local calculations 
on a lattice. 
The algorithm can conveniently divide the lattice and assign the sections
to different processors.  The communication between the nodes therefore 
is not extremely large.

We have tested the performance of our cluster in actual lattice QCD 
simulations.  Hioki and Nakamura \cite{hioki}
provide comparison performance data on SX-4 (NEC), SR2201(Hitachi),
 Cenju-3 (NEC) and  Paragon (Intel) machines.
Specifically, we compare the computing time per link update in microseconds 
per link and the inter-node communication speed in MB/sec. The link update is a
fundamental computational task within the QCD simulation and is therefore a 
useful standard.  The test was a simulation of improved pure guage lattice 
action ($1\times 1$ plaquet and $1\times 2$ rectangle terms) on a 
$16^4$ lattice. In each case the simulation was run on 16 processors.  We used
$\beta=6.0$

We used the QCDimMPI \cite{code} Fortran code.
Table \ref{table1} shows the results of this testing.

\begin{table}
\begin{center}
\caption{Comparison of performance of MPI QCD benchmark. Comparison data
from Hioki and Nakamura.}\label{table1}
\vspace{0.2cm}
\begin{tabular}{|c|c|c|} 
\hline 
\raisebox{0pt}[12pt][6pt]{Machine} & 
\raisebox{0pt}[12pt][6pt]{$\mu$-sec/link} & 
\raisebox{0pt}[12pt][6pt]{MB/sec}\\
\hline
\raisebox{0pt}[12pt][6pt]{SX-4} & 
\raisebox{0pt}[12pt][6pt]{4.50} & 
\raisebox{0pt}[12pt][6pt]{45}\\
\hline
\raisebox{0pt}[12pt][6pt]{SR2201} & 
\raisebox{0pt}[12pt][6pt]{31.4} & 
\raisebox{0pt}[12pt][6pt]{28}\\
\hline
\raisebox{0pt}[12pt][6pt]{Cenju-3} & 
\raisebox{0pt}[12pt][6pt]{57.42} & 
\raisebox{0pt}[12pt][6pt]{8.1}\\
\hline
\raisebox{0pt}[12pt][6pt]{Paragon} & 
\raisebox{0pt}[12pt][6pt]{149} & 
\raisebox{0pt}[12pt][6pt]{9.0}\\
\hline
\raisebox{0pt}[12pt][6pt]{\bf ZSU's Pentium cluster} & 
\raisebox{0pt}[12pt][6pt]{\bf 7.3} & 
\raisebox{0pt}[12pt][6pt]{\bf 11.5}\\
\hline

\end{tabular}
\end{center}
\end{table}
\vspace*{3pt}

\section{Conclusions}
We have reported on our efforts to build a parallel computing facility that
fits the demands and budget of a developing lattice field theory group.  We 
feel that a PC cluster can provide a very flexible and extremely economical
computing solution that is able to run parallel programs written using
the most popular message passing standard, MPI.  Furthermore we believe that 
this may in fact be the first such cluster at an academic physics institution 
in the People's Republic of China.

\section{Acknowledgements}
This work is supported by the
National Science Fund for Distinguished Young Scholars (19825117),
National Science Foundation, Guangdong Provincial Natural Science Foundation (990212) and 
Ministry of Education of China.
We are grateful for generous additional support from Guoxun 
(Guangdong National Communication Network) Ltd.. 
We would 
also like to thank Shinji Hioki of Tezukayama University for the use of the
QCDimMPI code.

\end{document}